# Adaptive Phase 2/3 Design with Dose Optimization

Cong Chen [a]*, Mo Huang [a], Xuekui Zhang [b]

[a] Biostatistics and Research Decision Sciences, Merck & Co., Inc., Rahway, NJ 07065, USA

*Corresponding author at: MAILSTOP UG-1CD44, 351 North Sumneytown Pike, North Wales, PA 19454, USA. E-mail address: cong_chen@merck.com

[b] Department of Mathematics and Statistics, University of Victoria, Canada

# Adaptive Phase 2/3 Design with Dose Optimization

## Abstract

FDA's Project Optimus initiative for oncology drug development emphasizes selecting a dose that optimizes both efficacy and safety. When an inferentially adaptive Phase 2/3 design with dose selection is implemented to comply with the initiative, the conventional inverse normal combination test is commonly used for Type I error control. However, indiscriminate application of this overly conservative test can lead to substantial increase in sample size and timeline delays, which undermines the appeal of the adaptive approach. This, in turn, frustrates drug developers regarding Project Optimus.

The inflation of Type I error depends on the probability of selecting a dose with better long-term efficacy outcome at end of the study based on limited follow-up data at dose selection. In this paper, we discuss the estimation of this probability and its impact on Type I error control in realistic settings. Incorporating it explicitly into the two methods we have proposed result in improved designs, potentially motivating drug developers to adhere more closely to an initiative that has the potential to revolutionize oncology drug development.

Keywords: combination test; early endpoint; overall survival; Project Endpoint

## 1. Introduction

The traditional paradigm for dose selection in oncology drug development originates from the experience with cytotoxic chemotherapeutics administered for a short period of time. Cytotoxic chemotherapy has a well-understood steep dose-response relationship, more dose equals better response, thus the chosen dose for Phase 2/3 development is often at or close to the maximum level tolerable by patients, aka the maximum tolerable dose (MTD). Novel therapies, on the other hand, are typically molecularly targeted and tend to have a plateau and saturation effect with diminishing response seen with additional drug. Moreover, targeted therapies are often given for much longer periods than cytotoxic chemotherapeutics, rendering the MTD approach less suitable.

To address this issue, the U.S. Food and Drug Administration (FDA) has recently launched Project Optimus, aiming to reform the dose optimization and selection paradigm [1]. The goal is to select a dose that not only optimizes efficacy but also safety and tolerability. This is a very worthy goal, however in practice there are two fundamental challenges that make this especially difficult to achieve. Firstly, when the dose selection is based on a limited number of patients with limited follow-up, which is often the case in practice, the selected dose may not be optimal among the candidate doses because the preliminary data used for decision making may not accurately predict long-term clinical benefits. Secondly, depending on the shapes of dose response curves for efficacy and tolerability, it may not be always possible to find a dose that can optimize efficacy and safety simultaneously. This is because a dose with superior efficacy may have inferior safety, and vice versa. Prioritizing efficacy over safety may lead to the selection of a dose that harms patients more than benefits. Conversely, prioritizing safety over efficacy may result in a dose that is associated with relatively better quality of life but may compromise the study's power on the primary efficacy endpoint.

There are three common oncology clinical development paths following the preliminary MTD finding (Fig. 1). Historically the lower pathway where one conducts a straightforward randomized Phase 3 study based upon the MTD has been the most common approach. However, with the advent of targeted therapies and the Project Optimus initiative, this approach has been less widely used. Instead, randomized dose-finding and adaptive 2/3 designs have become an integral part of the drug development paradigm. Under a typical (inferentially) adaptive Phase 2/3 design, multiple doses are included in Stage 1 (Phase 2) of the study and one dose is selected at end of Stage 1 to continue into Stage 2 (Phase 3). To control type I error and allow data to be used from both stages, an inverse normal combination test is often used for the primary analysis [2-3]. This approach appears promising as it not only provides randomized data for different doses but also provides randomized-controlled data for a solid proof-of-concept [4]. Nevertheless, the top pathway, a sequential approach of randomized dose optimization followed by a separate confirmatory Phase 3 study is more frequently taken in practice. One major concern about the adaptive approach amongst drug developers is the increased sample size and timeline delays due to the large statistical penalty paid under the conventional combination test. The test assumes that regardless of whether efficacy or safety is prioritized at dose selection, the dose with the best long-term efficacy outcome at the end of the study ("winner") is picked with absolute certainty, despite limited data. This assumption is apparently over conservative. Phase 3 oncology trial outcomes are notoriously difficult to predict from limited data, which is a major reason why the success rate is one of the lowest among all therapeutic areas.

The focus of our paper is on Type I error inflation and control of the adaptive Phase 2/3 design under a more reasonable assumption. It results in a more manageable increase of sample size, thus potentially making a promising alternative approach.

## 2. Multiplicity control of adaptive Phase 2/3 Design with dose selection

A hypothetical adaptive Phase 2/3 study with two candidate doses (low dose/high dose) will be used to facilitate the remaining discussion. The primary endpoint of the study is overall survival (OS), however there is no difference if any other long-term efficacy endpoint is used. The extent of Type I error inflation depends on how likely the selected dose has a better OS outcome than the unselected dose when there is no OS difference between the two (i.e., the probability of picking-the-winner due to random high). We assume that a dose will be selected at the time of analysis strictly following the pre-specified criteria. When the design allows the flexibility to carry both doses to Stage 2, Bonferroni-type full multiplicity adjustment may have to be applied.

Stage 1 patients treated at the unselected doses may have a different survival follow-up and treatment pattern compared to those treated at the selected dose, rendering the OS data unreliable for inclusion in the primary analysis. This may happen especially when the unselected dose has a detrimental effect on patients due to excessive toxicity. We assume that survival outcome for only the selected dose can be formally included in the primary analysis.

*Trial setup*

Our hypothetical study implements an equal randomization ratio in both stages. It includes two main analyses: a dose selection analysis shortly after all Stage 1 patients are enrolled, and a primary efficacy analysis at the end of the study after both Stage 1 and Stage 2 have sufficient follow-up for OS (Fig. 2). There is usually little OS information at the time of dose selection (e.g., <5% in the hypothetical study), which must rely on short-term efficacy and safety endpoints for the decision. However, Stage 1 patients contribute disproportionally more to the primary analysis due to longer follow-up in the end (e.g., 20% of OS information in the hypothetical study). Excluding the data from the primary analysis (e.g., under the sequential approach) makes a development program less efficient. In practice, an interim OS analysis may be conducted while Stage 2 is ongoing to potentially stop the study for overwhelming efficacy. Additionally, an interim analysis for potential accelerated approval may be conducted before Stage 2 enrollment is completed, following the one-trial approach under FDA's Project FrontRunner initiative [5].

*Probability of picking-the-winner (w)*

Suppose that dose selection at end of Stage 1 is based on both overall response rate (ORR) and Grade 3-4 adverse event (AE) rate. We consider two dose selection scenarios. In the first scenario, the lower dose will be selected only if the ORR is not lower by $C_x$ and the AE rate is at least lower by $C_s$ as compared to the higher dose. Otherwise, the higher dose will be selected. This scenario is biased towards the selection of higher dose because the lower dose must meet both criteria to be selected. This scenario is appropriate when the higher dose is preferred based on prior trial data and especially when there is flexibility of within-patient dose de-escalation if needed. In the second scenario, the higher dose will be selected only if the ORR is greater than the lower dose by $C_x$ and the AE rate is no greater than the lower dose by $C_s$. Otherwise, the lower dose will be selected. This scenario is biased towards the selection of lower dose.

Under the first scenario, the probability of picking-the-winner under the global null is

$$w=\Phi_3\left(0,\frac{C_x}{\sqrt{2R_x(1-R_x)/M}},-\frac{C_s}{\sqrt{2R_s(1-R_s)/M}};\rho_{xy},-\rho_{ys},-\rho_{xs}\right)$$
$$+0.5-\Phi_3\left(0,\frac{C_x}{\sqrt{2R_x(1-R_x)/M}},-\frac{C_s}{\sqrt{2R_s(1-R_s)/M}};-\rho_{xy},\rho_{ys},-\rho_{xs}\right)$$

Under the second scenario, the probability of picking-the-winner under the global null is

$$w=\Phi_3\left(0,-\frac{C_x}{\sqrt{2R_x(1-R_x)/M}},\frac{C_s}{\sqrt{2R_s(1-R_s)/M}};\rho_{xy},-\rho_{ys},-\rho_{xs}\right)$$
$$+0.5-\Phi_3\left(0,-\frac{C_x}{\sqrt{2R_x(1-R_x)/M}},\frac{C_s}{\sqrt{2R_s(1-R_s)/M}};-\rho_{xy},\rho_{ys},-\rho_{xs}\right)$$

where $\Phi_3$ is the cumulative joint distribution function of three standard normal variables, $\rho_{xs}$ is the correlation between ORR and the AE rate, $\rho_{xy}$ is the correlation between ORR and OS, $\rho_{ys}$ is the correlation between OS and the AE rate, $R_x$ is the average expected ORR between the two dose levels, $R_s$ is the average expected Grade 3-4 AE rate between the two dose levels, and $M$ is the sample size per group (=40 in the hypothetical study). See technical details in Appendix A.

For illustration purpose, we let $\rho_{xy} = 0.3$, reflecting a mild positive correlation between ORR and OS in Phase 3 trials, and $\rho_{xs} = 0.5$, reflecting the fact that tumor response and adverse event often increases with dose in Phase 1/2 trials. To conservatively assess the potential influence of safety and tolerability on OS, a concern within Project Optimus, three negative correlations $(-0.1,-0.3,-0.5)$ are assumed for $\rho_{ys}$ although Grade 3-4 AEs, when manageable, may not always have a negative impact on OS. Indeed, when the studied doses are all within a tolerable range with infrequent dose interruptions and discontinuations, a higher dose has often been preferred in oncology drug development, despite generally higher AE rates, a practice that has led to numerous drug approvals. Sometimes, certain AEs can even predict clinical benefit, such as skin rash in patients with non-small cell lung cancer after treatment with epidermal growth factor receptor tyrosine kinase inhibitors [6]. Furthermore, we set $C_s$ at 0.05, indicating that a difference in Grade 3-4 AE rate of less than 5% would not be considered clinically significant, and allowed $C_x$ to have a wide range from 0 to 0.2 to thoroughly assess the consequence of sacrificing early efficacy for safety in dose selection. $R_x$ and $R_s$ are less important and both set at 0.2.

Fig. 3 presents the probabilities of picking-the-winner under the above setup. As expected, for any fixed $C_x$ under each scenario, the probability increases as the negative impact of Grade 3-4 AE on OS increases. However, the impact overall is mild with the probabilities in the 50% to 65% range. While it generally increases with $C_x$ under scenario 1, it decreases with $C_x$ under scenario 2. In particular, the probability decreases to approximately 0.5 under scenario 2 when there is a >10% loss in ORR to improve the Grade 3-4 AE rate by 5%. Although $w$ is generally greater than 0.5 based on the above setup, it can be lower under different criteria for selecting the dose or parameter choices. For example, if we assume a positive correlation between Grade 3-4 AE and OS, choosing a dose with better safety would be like picking the loser in terms of OS. On the other hand, as the correlation between ORR and OS increases, sacrificing early efficacy may tend to decrease the probability of picking-the-winner. However, the overall trend in either scenario may not be monotone with correlation as it can be complicated by the correlation between Grade 3-4 AE and ORR, among other factors.

### *Impact of w on Type I error control*

We incorporate $w$ into methods, the exact parametric method and the exact inverse normal combination test, for Type I error control. Based on exact parametric calculation, the global null hypothesis on OS under the conventional log-rank test using the combined data from Stage 1 and Stage 2 patients (the selected dose vs control) should be tested at the $\alpha^E$ level that satisfies the following equation to control the overall Type I error at the target $\alpha$ level (see Appendix B for details):

$$(1 - \Phi_2(\Phi^{-1}(1 - \alpha^E); 1 - s/2))w + \Phi_2(\Phi^{-1}(\alpha^E); 1 - s/2)(1 - w) = \alpha$$

where $s$ is the information fraction of OS from Stage 1 patients, $\Phi$ is the cumulative distribution function of a standard normal variable and $\Phi^{-1}$ is its corresponding quantile, and $\Phi_2(x; \rho)$ is the cumulative joint distribution function at $(x, x)$ of two standard normal variables with correlation $\rho$. When $w = 0.5$, $\alpha^E = \alpha$ as the dose selection is totally random. When $w > 0.5$, $\alpha^E < \alpha$ because the tendency of picking-the-winner can inflate the Type I error; When $w < 0.5$, $\alpha^E > \alpha$ because the tendency of picking-the-loser can deflate the Type I error. To keep the Type I error under strong control, we let $\alpha^E = \alpha$ for $w \leq 0.5$.

Assuming that the pre-specified weight in the combination test is identical to the actual information fraction, the p-value for an exact inverse normal combination test is

$$p_c = 1 - \Phi\left(\sqrt{s}\Phi^{-1}(1 - p_{1a}) + \sqrt{1 - s}\Phi^{-1}(1 - p_{2s})\right)$$

where $p_{1a} = (1 - \Phi_2(\Phi^{-1}(1 - p_{1s}); 0.5))w + \Phi_2(\Phi^{-1}(p_{1s}); 0.5)(1 - w)$, and $p_{1s}$ and $p_{2s}$ are the respective p-values based on the two individual stages using the log-rank test. Like the exact parametric calculation, when $w = 0.5$ we have that $p_{1a} = p_{1s}$ as expected. To keep the Type I error under strong control, we may let $p_{1a} = p_{1s}$ for $w \leq 0.5$. The null hypothesis is rejected if $p_c < \alpha$. To measures the underlying efficiency of the exact combination test, we use the corresponding adjusted α-level (denoted by $\alpha^C$) to the log-rank test. The efficiencies of the conventional combination tests under $w = 1$ using the Dunnett and Sidak adjustments for Stage 1 dose selection [7] can be measured similarly.

Fig. 4 adjusted α-levels under the two exact methods ($\alpha^E$ and $\alpha^C$) as well as the adjusted α-levels based on the Dunnett and Sidak adjustments. Despite difference in the underlying mathematics, the two exact methods yield almost identical results. The same findings are observed at different *s* (results not presented here). When *w* is in the range of 50% to 65%, they outperform the Dunnett and Sidak adjustments by a wide margin. With the corresponding $\alpha^E$ and $\alpha^C$ greater than 2.2%, the sample size increase of the adaptive design approach due to multiplicity adjustment becomes more acceptable to drug developers. As a result, more may be encouraged to take the approach despite its operationally complexity.

*Estimation of the correlation between two test statistics*

To estimate the correlations in the above assessment, note that since test statistics are normalized estimates of treatment effects, the correlation between two test statistics is the same as for the corresponding treatment effect estimates. This correlation is largely determined by the individual-level correlation of the respective endpoints and should not be confused with trial-level correlation, which plays a key role in statistical design of proof-of-concept trials [8].

At the design stage, the correlations involved in the above Type I error assessment should be carefully determined from historical trials with similar design (e.g., in terms of allowance of cross-over, data maturity at final analysis) for drugs in similar classes. The estimation of correlation is relatively straightforward when the two endpoints are either binary or continuous, whereas single-arm trials can be used to assist with the estimation. In general, data from randomized-controlled trials must be used for more reliable estimation especially if a time-to-event endpoint is involved. When patient-level data from a historical trial is available, a bootstrapping method may be used for the estimation. Each bootstrap sample should be representative of the overall patient population in terms of follow-up patterns and be balanced with the key prognostic factors between the two treatment arms. A Pearson correlation estimate for two test statistics is obtained after the corresponding treatment effects across the samples are estimated [9]. When subgroup analyses from a published historical trial are available, a modified Pearson correlation estimate may be considered (see Appendix C for details). In practice, estimates from different data sources can be pooled to provide a more robust estimate as well as a reasonable range.

Ideally, to avoid complexity, the estimates of correlations from historical trials are directly incorporated into not only the planning but also the analysis of a prospective study. When the current trial data must be used for the estimation of correlations, details of the method (e.g., bootstrapping) should be pre-specified. Though not a common practice in the industry, this approach is not much dissimilar in spirit from the well-accepted group sequential design method, where correlation is also estimated, or from the multiple imputation method routinely applied to the analysis of longitudinal trial data. In our hypothetical example, to be conservative, the individual-level correlation between AE and OS is assumed to be negative. It may go either direction: potentially negative when the doses are less tolerable and potentially positive otherwise, and the correlation may be weaker than assumed in the hypothetical design. As the standard-of-care improves and more effective therapies become available, the individual-level correlation between ORR and OS in earlier line settings may become weaker. If the correlations for the endpoints involved in dose selection are deemed small or their impact associated with the dose selection criteria on Type I error is negligible, no penalty may be paid. To maintain consistency with conventional practice, an administrative alpha will be allocated during dose selection to prevent premature efficacy stopping, effectively paying a small penalty. The precedence of paying no penalty has been successfully set in the adaptive Phase 2/3 study of Gardasil, an HPV vaccine for preventing certain strains of human papillomavirus [10].

## 3. Discussions

The emphasis of Project Optimus on safety and tolerability may result in a dose (usually the lower one) with relatively better quality of life, but not necessarily improved OS. In this paper, we have demonstrated that in realistic settings with two candidate doses, the probability of the selected dose yielding a better OS outcome than the unselected dose due to random high is only ~50-65%. Incorporating it explicitly into the two exact methods we have proposed result in improved designs than under the conventional approach, potentially motivating drug developers to adhere more closely to an initiative that has the potential to revolutionize oncology drug development. Further comparison between the two methods, as well as their comparison with operational adaptive Phase 2/3 designs in terms of practical considerations, is part of ongoing investigation.

The proposed exact methods can be extended to more than two doses after additional mathematical derivations, although it is uncommon to include more than two in a registrational oncology study. Throughout the paper, we have implicitly assumed that two doses have different safety and efficacy profiles. However, when the doses are located at or close to the plateau of a dose–response curve, one may take the "MiniPool" approach by pooling the data from the less effective dose in Stage 1 and the selected dose in Stage 2 for the primary analysis without paying any penalty [11]. This approach avoids the complexity of multiplicity adjustment and is efficient when the treatment effect between the two doses is expected to be small. When there are multiple efficacy and safety endpoints in dose selection, we may prioritize by selecting a representative efficacy endpoint and a safety endpoint, and then follow the approaches outlined in this paper to streamline the statistical evaluation. Alternatively, an extensive simulation study may be conducted for Type I error evaluation, following relevant regulatory guidance on adaptive designs.

While this work is motivated by Project Optimus, adaptive Phase 2/3 design can also play a crucial role in Project Endpoint and Project FrontRunner, two other closely related initiatives launched by the FDA in recent years. When an investigational drug potentially induces significant toxicity, substantial treatment effect in ORR (or other efficacy endpoints) may not lead to OS benefit [12]. When drug toxicity is in question, Project Endpoint requests additional survival data to better characterize the drug's risk-benefit profile before any approval, even if OS is not the primary endpoint [13]. Similarly, a substantial ORR effect at the interim analysis of a Phase 3 study may not be sufficient to justify the accelerated approval under the hallmark one-trial approach of Project FrontRunner. The adaptive Phase 2/3 design, inferentially or operationally, potentially alleviates the concern. With dose optimization based on randomized comparison, there will be greater certainty about the risk-benefit profile of the selected dose. Less OS data may be required, and faster approval may be achieved.

**Acknowledgement**

During the preparation of this manuscript, we have received numerous comments and technical assistance from Merck colleagues, the members of cross-industry 2-in-1 design team, as well as multiple current and former FDA statisticians. We have also benefited from the discussions at a recent FDA-AACR Dosing Workshop, where the statistical issue with the current approach was presented.

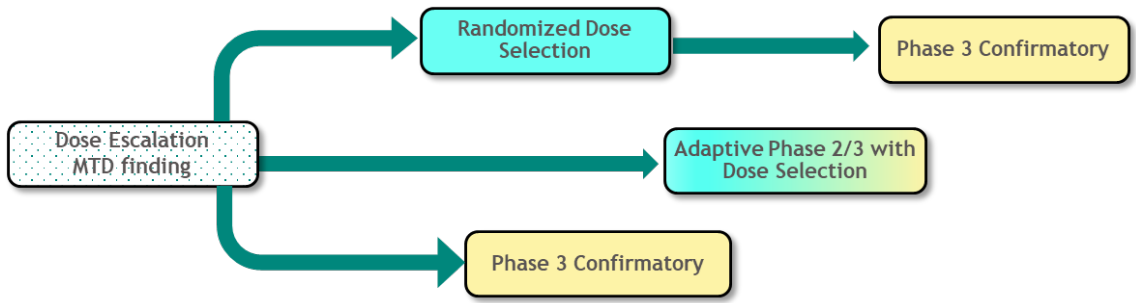

Fig. 1. Three typical oncology clinical development paths post preliminary dose finding: 1) randomized dose finding followed with a confirmatory Phase 3 study (top); 2) inferentially adaptive Phase 2/3 with dose selection (middle); 3) straight confirmatory Phase 3 study (bottom).

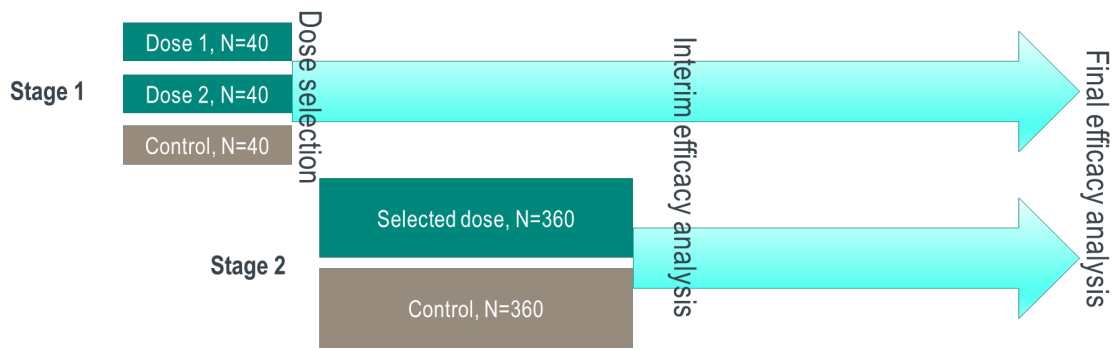

Fig. 2. A hypothetical inferentially adaptive Phase 2/3 study with dose selection comprises three analyses: 1) dose selection based on both efficacy and safety; 2) interim analysis of the primary efficacy endpoint; 3) final analysis of the primary efficacy endpoint. An additional interim analysis of surrogate endpoints may be conducted in practice for potential accelerated approval before Stage 2 enrollment is completed.

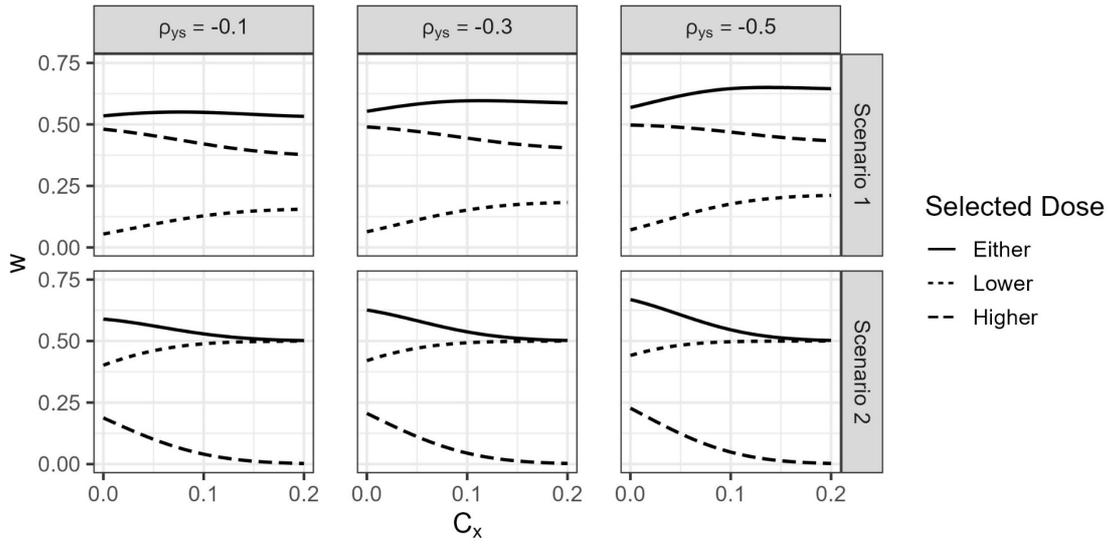

Fig. 3. Probability of picking-the-winner under the two scenarios (top: scenario 1; bottom: scenario 2) for different correlations between Grade 3-4 AE and OS (-0.1, -0.3, -0.5) when the bar for ORR difference ranges from 0 to 20%.

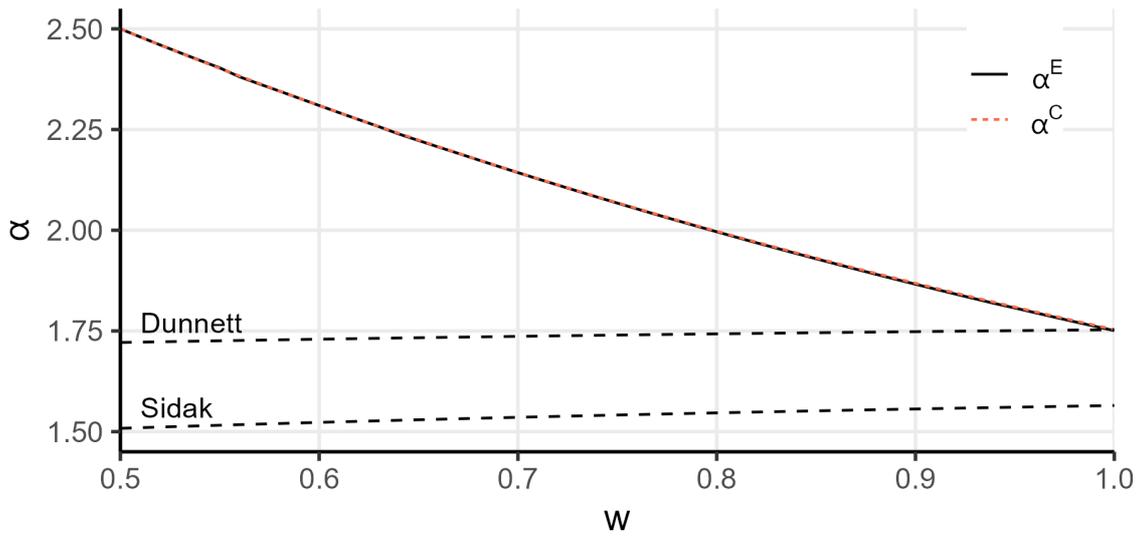

Fig. 4. The adjusted α-levels under the two exact methods associated with the probability of picking-the-winner (the two lines overlap), as well as the Dunnett and Sidak adjustments of the inverse normal combination test.

**Appendix A. Details on calculation of *w* under the global null**

Consider a two-stage randomized controlled inferentially adaptive Phase 2/3 design that starts with two dose levels in Stage 1 and carries one dose into Stage 2. The OS data from the two stages are combined for the primary analysis. Let $Y_{1j}$ be the log-rank test of OS in Stage 1 patients at dose level $j$ in the primary analysis, $j = 1, 2$, and let $Y_{1s}$ be the corresponding log-rank test for the selected dose. Let $(X_1, X_2)$ be the ORRs and $(S_1, S_2)$ be the AE rates in the two dose groups, each with $M$ patients. Under the null hypothesis, $X_1 - X_2$ has an approximate $N(0, 2R_x(1 - R_x)/M)$ distribution where $R_x$ is the average ORR between the two dose groups and $S_1 - S_2$ has an approximate $N(0, 2R_s(1 - R_s)/M)$ distribution where $R_s$ is the average AE rate between the two dose groups. Let $w = P(Y_{1s} = \max\{Y_{11}, Y_{12}\})$, the probability of picking a winner.

*Estimation of w*

In the first scenario, dose 1 will be selected if the ORR is not lower by $C_x$ and the AE rate is at least lower by $C_s$ as compared to dose 2. Otherwise, dose 2 will be selected. The probability of dose 1 being the winner is

$$w_1 = P(Y_{11} - Y_{12} > 0 | X_1 - X_2 > -C_x, S_1 - S_2 < -C_s) P(X_1 - X_2 > -C_x, S_1 - S_2 < -C_s)$$
$$= P(Y_{11} - Y_{12} > 0, X_1 - X_2 > -C_x, S_1 - S_2 < -C_s)$$
$$= P(Y_{12} - Y_{11} < 0, X_2 - X_1 < C_x, S_1 - S_2 < -C_s)$$
$$= \Phi_3\left(0, \frac{C_x}{\sqrt{2R_x(1-R_x)/M}}, -\frac{C_s}{\sqrt{2R_s(1-R_s)/M}}; \rho_{xy}, -\rho_{ys}, -\rho_{xs}\right)$$

Similarly, the probability of dose 2 being the winner is

$$w_2 = P(Y_{12} - Y_{11} > 0, X_1 - X_2 < -C_x \text{ or } S_1 - S_2 > -C_s)$$
$$= P(Y_{12} - Y_{11} > 0) - P(Y_{12} - Y_{11} > 0, X_1 - X_2 > -C_x, S_1 - S_2 < -C_s)$$
$$= 0.5 - P(Y_{11} - Y_{12} < 0, X_2 - X_1 < C_x, S_1 - S_2 < -C_s)$$
$$= 0.5 - \Phi_3\left(0, \frac{C_x}{\sqrt{2R_x(1-R_x)/M}}, -\frac{C_s}{\sqrt{2R_s(1-R_s)/M}}; -\rho_{xy}, \rho_{ys}, -\rho_{xs}\right)$$

where $\Phi_3$ is the cumulative joint distribution function of three standard normal variables.

In the second scenario, dose 1 will be selected if the ORR is higher than dose 2 by $C_x$ and the AE rate is no worse than dose 2 by $C_s$. Otherwise, dose 2 will be selected. The probability of dose 1 being the winner is

$$w_1 = P(Y_{11} - Y_{12} > 0, X_1 - X_2 > C_x, S_1 - S_2 < C_s)$$
$$= P(Y_{12} - Y_{11} < 0, X_2 - X_1 < -C_x, S_1 - S_2 < C_s)$$
$$= \Phi_3\left(0, -\frac{C_x}{\sqrt{2R_x(1-R_x)/M}}, \frac{C_s}{\sqrt{2R_s(1-R_s)/M}}; \rho_{xy}, -\rho_{ys}, -\rho_{xs}\right)$$

and the pprobability of dose 2 being the winner is

$$w_2 = P(Y_{12} - Y_{11} > 0, X_1 - X_2 < C_x \text{ or } S_1 - S_2 > C_s)$$
$$= P(Y_{12} - Y_{11} > 0) - P(Y_{12} - Y_{11} > 0, X_1 - X_2 > C_x, S_1 - S_2 < C_s)$$
$$= 0.5 - P(Y_{11} - Y_{12} < 0, X_2 - X_1 < -C_x, S_1 - S_2 < C_s)$$
$$= 0.5 - \Phi_3\left(0, -\frac{C_x}{\sqrt{2R_x(1-R_x)/M}}, \frac{C_s}{\sqrt{2R_s(1-R_s)/M}}; -\rho_{xy}, \rho_{ys}, -\rho_{xs}\right)$$

Under either scenario, the overall probability of picking-the-winner is $w = w_1 + w_2$.

See R code below for generating $w_1$ under the first dose selection scenario.

```
calc_w1 <- function(Cx, Cs, rxy, rys, rxs, M, Rx, Rs) {
  sig <- matrix(c(1, rxy, -rys, rxy, 1, -rxs, -rys, -rxs, 1), 3, 3)
  yubound <- 0
  xubound <- Cx/sqrt(2*Rx*(1-Rx)/M)
  subound <- -Cs/sqrt(2*Rs*(1-Rs)/M)
  pmvnorm(lower = c(-Inf, -Inf, -Inf),
      upper = c(yubound, xubound, subound),
      sigma = sig)[1]
}
```

**Appendix B. The impact of $w$ on Type I error control**

At the time of dose selection, patients have limited follow-up and there is great uncertainty about whether the selected dose has a larger OS effect than the unselected dose. Denote by $Y_{2s}$ the log-rank test of OS based on Stage 2 patients in the primary analysis. The conventional log-rank test for the combined data can be approximated with $\sqrt{s}Y_{1s} + \sqrt{1-s}Y_{2s}$, where $s$ is the information fraction of OS from Stage 1 patients. However, $Y_{1s}$ does not have a standard normal distribution due to dose selection, which complicates the statistical characterization of $\sqrt{s}Y_{1s} + \sqrt{1-s}Y_{2s}$.

The randomization ratio of the study is assumed to be $r$:1 between a dose level and control, i.e., $r$:$r$:1 (dose 1 vs dose 2 vs control) in Stage 1 and $r$:1 (selected dose vs control) in Stage 2. Due to the sharing of control, we have that $corr(Y_{11}, Y_{12}) = 1/(1+r)$. Since Stage 1 and Stage 2 patients are independent, we also have that $corr(Y_{1j}, Y_{2s}) = 0$ ($j$=1,2). As a result, $corr(\sqrt{s}Y_{11} + \sqrt{1-s}Y_{2s}, \sqrt{s}Y_{12} + \sqrt{1-s}Y_{2s}) = s/(1+r) + 1 - s$.

*Exact parametric calculation*

The null hypothesis will be tested at an adjusted $\alpha^E$ level based on $\sqrt{s}Y_{1s} + \sqrt{1-s}Y_{2s}$. the overall Type I error rate under weak control is

$$\sum_{j=1}^{2} P(\sqrt{s}Y_{1s} + \sqrt{1-s}Y_{2s} > \Phi^{-1}(1-\alpha^E)|Y_{1s} = Y_{1(j)}) P(Y_{1s} = Y_{1(j)})$$

$= P(\sqrt{s}\max\{Y_{11}, Y_{12}\} + \sqrt{1-s}Y_{2s} > \Phi^{-1}(1-\alpha^E))w$
$\quad + P(\sqrt{s}\min\{Y_{11}, Y_{12}\} + \sqrt{1-s}Y_{2s} > \Phi^{-1}(1-\alpha^E))(1-w)$

$= \{1 - P(\sqrt{s}Y_{11} + \sqrt{1-s}Y_{2s} < \Phi^{-1}(1-\alpha^E), \sqrt{s}Y_{12} + \sqrt{1-s}Y_{2s} < \Phi^{-1}(1-\alpha^E))\}w$
$\quad + P(\sqrt{s}Y_{11} + \sqrt{1-s}Y_{2s} > \Phi^{-1}(1-\alpha^E), \sqrt{s}Y_{12} + \sqrt{1-s}Y_{2s} > \Phi^{-1}(1-\alpha^E))(1-w)$

$= \{1 - P(\sqrt{s}Y_{11} + \sqrt{1-s}Y_{2s} < \Phi^{-1}(1-\alpha^E), \sqrt{s}Y_{12} + \sqrt{1-s}Y_{2s} < \Phi^{-1}(1-\alpha^E))\}w$
$\quad + P(\sqrt{s}Y_{11} + \sqrt{1-s}Y_{2s} > \Phi^{-1}(1-\alpha^E), \sqrt{s}Y_{12} + \sqrt{1-s}Y_{2s} > \Phi^{-1}(1-\alpha^E))(1-w)$

$= \{1 - P(\sqrt{s}Y_{11} + \sqrt{1-s}Y_{2s} < \Phi^{-1}(1-\alpha^E), \sqrt{s}Y_{12} + \sqrt{1-s}Y_{2s} < \Phi^{-1}(1-\alpha^E))\}w$
$\quad + P(\sqrt{s}Y_{11} + \sqrt{1-s}Y_{2s} < \Phi^{-1}(\alpha^E), \sqrt{s}Y_{12} + \sqrt{1-s}Y_{2s} < \Phi^{-1}(\alpha^E))(1-w)$

$= \left(1 - \Phi_2(\Phi^{-1}(1-\alpha^E); \frac{s}{1+r} + 1 - s)\right)w + \Phi_2\left(\Phi^{-1}(\alpha^E); \frac{s}{1+r} + 1 - s\right)(1-w)$

Setting the above equation at the desired $\alpha$ level for the study, $\alpha^E$ can be easily obtained (See R code below). When $w = 0.5$, given that $P(\sqrt{s}\max\{Y_{11}, Y_{12}\} + \sqrt{1-s}Y_{2s} > c) + P(\sqrt{s}\min\{Y_{11}, Y_{12}\} + \sqrt{1-s}Y_{2s} > c) = P(\sqrt{s}Y_{11} + \sqrt{1-s}Y_{2s} > c) + P(\sqrt{s}Y_{12} + \sqrt{1-s}Y_{2s} > c)$ for any $c$, the first equality implies that $\alpha^E = \alpha$ as expected. Since $\alpha^E$ clearly decreases as $w$ increases, the Type I error can be kept under strong control (i.e., there is no treatment effect at the selected dose but there may be treatment effect at the unselected dose) by letting $\alpha^E = \alpha$ for $w \leq 0.5$.

```
winner_p <- function(alphaE, s, r) {
q <- qnorm(1 - alphaE)
1 - pmvnorm(upper = c(q, q),
sigma = matrix(c(1, s/(1+r) + 1 - s, s/(1+r) + 1 - s, 1), 2, 2),
keepAttr = FALSE)
}
loser_p <- function(alphaE, s, r) {
q <- qnorm(alphaE)
pmvnorm(upper = c(q, q),
sigma = matrix(c(1, s/(1+r) + 1 - s, s/(1+r) + 1 - s, 1), 2, 2),
keepAttr = FALSE)
}
calc_alpha <- function(alphaE, s, r, w) {
winner_p(alphaE, s, r)*w + loser_p(alphaE, s, r)*(1-w)
}
calc_alphaE <- function(alpha, s, r, w) {
uniroot_fn <- function(alphaE) {
calc_alpha(alphaE, s, r, w) - alpha
}
uniroot(uniroot_fn, interval = c(1e-6, 0.1))$root
}
```

### *Exact combination test*

Let $p_{1s} = 1 - \Phi^{-1}(Y_{1s})$ be the Stage 1 p-value based on observed $Y_{1s}$, and $p_{2s} = 1 - \Phi^{-1}(Y_{2s})$ be the Stage 2 p-value based on observed $Y_{2s}$. Following the same derivation as above, the adjusted p-value of Stage 1 after dose selection is

$$p_{1a} = P(\max\{Y_{11}, Y_{12}\} > \Phi^{-1}(1 - p_{1s}))w + P(\min\{Y_{11}, Y_{12}\} > \Phi^{-1}(1 - p_{1s}))(1 - w)$$
$$= \left(1 - \Phi_2(\Phi^{-1}(1 - p_{1s}); \frac{1}{1+r})\right)w + \Phi_2\left(\Phi^{-1}(p_{1s}); \frac{1}{1+r}\right)(1 - w)$$

Assuming that the pre-specified weight is identical to the actual information fraction, the p-value of the weighted inverse normal combination test is

$$p_c = 1 - \Phi\left(\sqrt{s}\Phi^{-1}(1 - p_{1a}) + \sqrt{1-s}\Phi^{-1}(1 - p_{2s})\right)$$

Like exact calculation, when $w = 0.5$ we have that $p_{1a} = p_{1s}$ as expected. To keep the Type I error under strong control, we may let $p_{1a} = p_{1s}$ for $w \leq 0.5$.

The null hypothesis is rejected if $p_c < \alpha$. Let $\alpha^C = 1 - \Phi\left(\sqrt{s}\Phi^{-1}(1 - p_{1s}) + \sqrt{1-s}\Phi^{-1}(1 - p_{2s})\right)$ be the corresponding adjusted α-level to the log-rank test. For fixed $w$, $r$, and $s$, an approximate value of $\alpha^C$ is obtained after averaging it over the joint distribution of $(p_{1s}, p_{2s})$ under the constraint $p_c = \alpha$. Specifically, $p_{2s}$ is first sampled from a uniform distribution. The corresponding value of $p_{1a}$ is calculated such that $p_c = \alpha$. $p_{1s}$ is calculated from each $p_{1a}$ by applying the inverse adjustment function. The density function of $Y_{1s} = 1 - \Phi^{-1}(p_{1s})$ is $f(Y_{1s}) = w * f(\max(Y_{11}, Y_{12})) + (1 - w) * f(\min(Y_{11}, Y_{12}))$. $\alpha^C$ is then calculated as a weighted average of $\sqrt{s}\Phi^{-1}(1 - p_{1s}) + \sqrt{1-s}\Phi^{-1}$ weighted by $f(Y_{1s})$ and transformed back to the p-value scale. We can similarly calculate the corresponding adjusted α-levels when the Sidak adjustment $1 - (1 - p_{1s})^2$ and the Dunnett adjustment $P(\max\{Y_{11}, Y_{12}\} > \Phi^{-1}(1 - p_{1s}))$ are used for Stage 1 dose selection. See R code example below.

```
### Exact
###The exact distribution of max/min of two correlated normal variables can be found in
https://gwern.net/doc/psychology/personality/conscientiousness/2008-nadarajah.pdf

library(tidyverse)
library(mvtnorm)

alpha <- 0.025
s <- 0.2
p2samp <- ppoints(1e4)
p1samp <- 1-pnorm((qnorm(1-alpha)-sqrt(1-s)*qnorm(1-p2samp))/sqrt(s))

ind <- p1samp > 1e-6 & p2samp > 1e-6

p1samp <- p1samp[ind]
p2samp <- p2samp[ind]

inv_p1a <- function(p1a, w) {
  opt_fn <- function(p1s, w) {
    (calc_p1a(p1s, w) - p1a)^2
  }
  optimize(opt_fn, interval = c(p1a, inv_sidak(p1a)), w = w,
        tol = 1e-8)$minimum
}

exact_c <- rep(0, length(wrange))
for (i in seq_along(wrange)) {
  w <- wrange[i]
  out <- tibble(p1a = p1samp, p2 = p2samp) %>%
    rowwise() %>%
```

```
    mutate(p1s = inv_p1a(p1a, w = w)) %>%
    ungroup() %>%
    mutate(Zc = qnorm(1-combination_test(p1s, p2, s = s)),
        q1 = qnorm(1-p1s),
        d = 2*dnorm(q1)*pnorm(0.5*q1/sqrt(3/4))*w +
          2*dnorm(q1)*pnorm(-0.5*q1/sqrt(3/4))*(1-w)) %>%
    filter(d != 0)
  exact_c[i] <- 1-pnorm(sum(out$Zc*out$d)/sum(out$d))
  print(w)
}
```

**Appendix C. Correlation estimation based on published subgroup analyses**

The treatment effect estimates across the baseline variables are highly correlated, rendering the naïve Pearson correlation estimate inappropriate. However, despite the small number of subgroups under each baseline variable (usually 2-3), unbiased estimates of variances for the two treatment effects and covariance between the two can be obtained under each baseline variable due to the independence of the subgroups. By averaging over the baseline variables, we obtain the mean estimates of the covariance and variances, leading to a reasonable estimate of the correlation. In the special case of two subgroups per baseline variable, the modified Pearson correlation estimate takes the form of $\frac{\sum(T_{1i1}-T_{1i2})(T_{2i1}-T_{2i2})}{\sqrt{\sum(T_{1i1}-T_{1i2})^2 \sum(T_{2i1}-T_{2i2})^2}}$ where $T_{lij}$ is the estimated treatment effect on endpoint $l$ for the $j$-th subgroup under the $i$-th baseline variable by noticing that each summand in the numerator is an unbiased estimate of the covariance and each summand in the denominator is an unbiased estimate of the variance. When there are more than two subgroups under a baseline variable, one may collapse them into two to directly apply this formular or follow the above general approach to arrive at a reasonable estimate.

Figure A presents the scatter plot of negative log(hazard ratio) of OS and ORR difference based on EV-302, a Phase 3 study of enfortumab vedotin (an antibody drug conjugate) and pembrolizumab (an anti-PD-1 immunotherapy) vs chemotherapy in untreated advanced urothelial cancer (DOI: 10.1056/NEJMoa2312117). In this example, the estimated correlation between the two is 0.32. This may serve as a starting point for assessing the correlation of test statistics between ORR and OS in a similarly designed Phase 3 study that tests the combination of an antibody drug conjugate and an anti-PD-(L)1 immunotherapy.

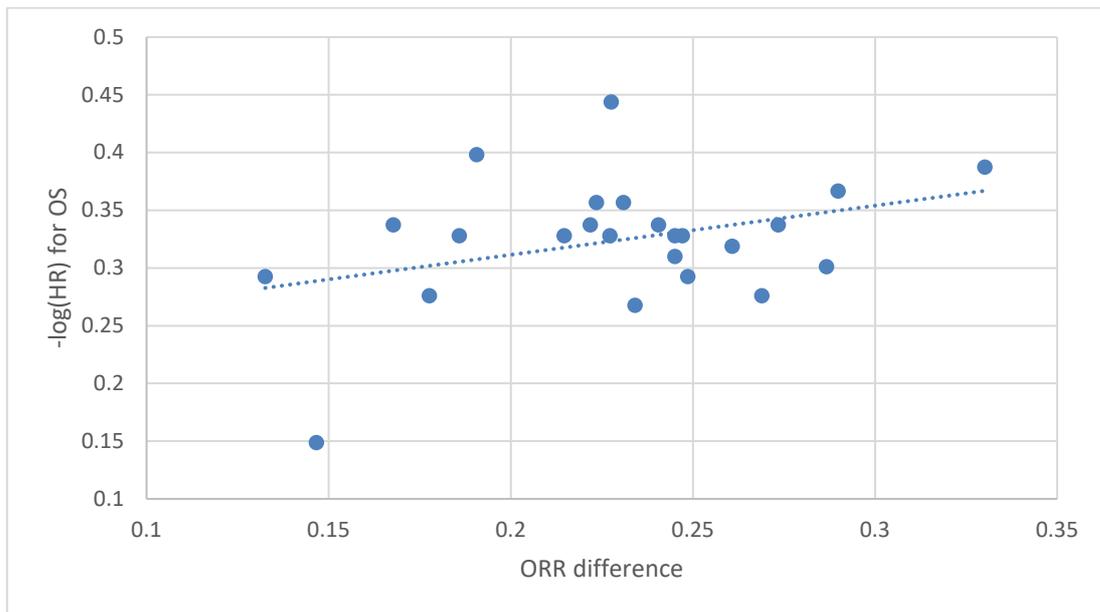

Fig. A. Scatter plot of -log(HR) of OS and ORR difference and the linear regression line based on published subgroup data from the Phase 3 EV-302 study in untreated advanced urothelial cancer. The Pearson correlation implied in the scatter is not an appropriate estimate of the true correlation due to the correlations among subgroups.